\documentclass{article}

\usepackage[nonatbib,final]{neurips_2020}
\usepackage[numbers]{natbib}
\usepackage[utf8]{inputenc} % allow utf-8 input
\usepackage[T1]{fontenc}    % use 8-bit T1 fonts
\usepackage{hyperref}       % hyperlinks
\usepackage{url}            % simple URL typesetting
\usepackage{booktabs}       % professional-quality tables
\usepackage{amsfonts}       % blackboard math symbols
\usepackage{nicefrac}       % compact symbols for 1/2, etc.
\usepackage{microtype}      % microtypography
\usepackage{mathtools}
\usepackage{subcaption}
\usepackage{graphicx}
\usepackage{amsthm}
\usepackage[noabbrev,capitalize]{cleveref}
\usepackage{xcolor}
\usepackage{xspace}
\usepackage{algorithm}
\usepackage{algpseudocode}

\newtheorem{theorem}{Theorem}
\newtheorem{lema}{Theorem}
\newtheorem{assum}{Theorem}
\newtheorem{defi}{Theorem}
\newtheorem{definition}[defi]{Definition}
\newtheorem{lemma}[lema]{Lemma}
\newtheorem{assumption}[assum]{Assumption}

\theoremstyle{remark}

\usepackage{titletoc}

\usepackage{comment}
\definecolor{mygray}{gray}{0.6}

\newcommand{\rank}{\mathbf{rank}}
\newcommand{\name}{DCF-PCA\xspace}

\title{Distributed Robust Principal Component Analysis}

\author{%
  Wenda Chu\\
  Institute for Interdisciplinary Information Sciences\\
  Tsinghua University\\
  \texttt{chuwd19@mails.tsinghua.edu.cn}
}

\begin{document}

\maketitle

\begin{abstract}
We study the robust principal component analysis (RPCA) problem in a distributed setting. The goal of RPCA is to find an underlying low-rank estimation for a raw data matrix when the data matrix is subject to the corruption of gross sparse errors. Previous studies have developed RPCA algorithms that provide stable solutions with fast convergence. However, these algorithms are typically hard to scale and cannot be implemented distributedly, due to the use of either SVD or large matrix multiplication. In this paper, we propose the first distributed robust principal analysis algorithm based on consensus factorization, dubbed DCF-PCA. We prove the convergence of DCF-PCA and evaluate DCF-PCA on various problem settings.

\end{abstract}
\section{Introduction}
\label{sec:intro}

Principal robust analysis (PCA) has been widely used for dimension reduction in data science. It extracts the top $k$ significant components of a given matrix by computing the best low-rank approximation. However, it is well known that PCA is sensitive to noises and adversarial attacks. Robust PCA (RPCA) aims at mitigating this drawback by separating the noise out explicitly. Specifically, RPCA assumes that the observed matrix $M\in \mathbb R^{m\times n}$ can be decomposed as $M = L^* + S^*$ where $L^*$ is a low-rank matrix and $S^*$ is a sparse matrix. The goal of RPCA is to recover the low-rank matrix $L^*$ from the noisy data $M$, which is typically expressed as an optimization problem:

\begin{equation}
\label{eq:origin_problem}
    \min_{L,S} \rank (L) + \lambda \|S\|_0, \ \mathrm{s.t.}\ M = L + S.
\end{equation}

Unfortunately, this optimization problem is known to be NP-hard. Therefore, \cref{eq:origin_problem} is often reformulated to other optimization problems listed below.

\begin{itemize}
    \item Convex relaxation of $\rank(\cdot)$ to nuclear norm $\|\cdot\|_*$ and $\ell_0$ norm to $\ell_1$ norm.
    \begin{equation}
    \label{eq:convex1}
        \min_{L,S} \|L\|_* + \lambda \|S\|_1,\ \mathrm{s.t.}\ M = L+S.
    \end{equation}
    where the nuclear norm is defined as sum of singular values $\|L\|_* = \sum_{i=1}^{\min(m,n)} \sigma_i(L)$ and $\|S\|_1$ denotes the $\ell_1$ norm of $S$ as a vector: $\|S\|_1 = \sum_{ij}|S_{ij}|$.
    
    \item A variant of RPCA considers recovering $L$ from another noise with bounded Frobenius norm. The corresponding optimization problem is formulated as
    \begin{equation}
    \label{eq:convex2}
        \min_{L,S} \|L\|_* + \lambda \|S\|_1 + \frac{\mu}{2}\|L + S - M\|_F^2.
    \end{equation}
    
    \item Other works attempt to solve RPCA based on the low-rank matrix factorization that decomposes a low-rank matrix $M$ as $M = UV^T$ ($U\in \mathbb R^{m\times r}, V\in \mathbb R^{n\times r}$). The rank function is resolved by an implicit constraint that $\rank(UV^T)\leq r$. Feng el al. \cite{Feng2013OnlineRP} proposed to optimize a nonconvex problem:
    \begin{equation}
    \label{eq:noncovex_problem}
        \min_{U,V,S} \frac{1}{2} \|UV^T + S - M\|^2_F + \frac{\rho}{2}(\|U\|^2_F + \|V\|^2_F) + \lambda\|S\|_1
    \end{equation}
    which exploited the property of nuclear norm \cite{Recht_2010} so that its global minimum also minimizes the objective of \cref{eq:convex2}:
    \begin{equation}
        \|L\|_* = \inf_{U,V} \left\{\frac{1}{2}\|U\|_F^2 + \frac{1}{2}\|V\|_F^2: UV^T = L\right\}
    \end{equation}
\end{itemize}

The convex problems \cref{eq:convex1,eq:convex2} are well-posed and can be optimized using standard convex optimization methods, e.g., SDP, PGM and ADMM. However, the existence of the nuclear norm $\|\cdot\|_*$ makes it difficult for these algorithms to scale, since it cannot be calculated distributively.

In this paper, we present a distributed RPCA algorithm based on consensus factorization (\name), which solves the nonconvex problem \cref{eq:noncovex_problem} distributedly. We formalize the distributed RPCA problem and elaborate our \name algorithm in \cref{sec:2}; provides theoretical guarantees for \name in \cref{sec:theoretical}; and exhibits numerical results for \name in \cref{sec:exp}.

\section{Distributed Robust Principal Component Analysis}
\label{sec:2}
\subsection{Problem Definition}
\label{subsec:DRPCA}
We formalize the problem of distributed robustness principal component problem. Assume the data $M\in \mathbb R^{m\times n}$ are distributed over $E$ remote clients. Each client $i\in [E]$ only has access to some columns of the matrix $M$.
\begin{equation}
\setlength\arraycolsep{2pt}
    M = \begingroup
    \begin{bmatrix}
    M_1 & M_2 & \dots & M_E\end{bmatrix}
    \endgroup,\ \text{where}\ M_i \in \mathbb R^{m\times n_i}\ \text{and}\ n = \sum_{i=1}^E n_i.
\end{equation}
For clarity, we define $L_i^*$ and $S_i^*$ such that $M_i = L_i^* + S_i^*$ for all $i\in [E]$.

\textbf{Limited Communication.} As the communication cost over remote devices are typically high, we give limited communication budget for the clients. Assuming $m = O(n)$, naively broadcasting the whole matrix needs transmitting a prohibitively large amount of data $O(n^2 K)$.

\textbf{Privacy Preserving.} Privacy is considered as one of the most crucial issues in distributed learning. In practice, some local data $M_i$ may be privacy-sensitive and cannot be exposed to other clients. We call a distributed scheme \emph{privacy-preserving} for a set of sensitive clients $\mathcal I$, if it recovers $L_i$ for $i\not\in \mathcal I$ but protects $M_i$ for $i\in \mathcal I$.

\subsection{Distributed RPCA algorithm via consensus factorization}
\label{subsec:DRPCA_algo}

\begin{algorithm}[tb]
\caption{Distributed RPCA algorithm via consensus factorization (\name)}
\label{algorithm1}
\textbf{Input:} $E$ remote clients with submatrices $\{M_1,\dots, M_E\}$.
\begin{algorithmic}
\State Initialize matrix $U^{(0)}\in \mathbb R^{m\times r}$. Each client $i\in [E]$ randomly initializes $V_i\in \mathbb R^{n_i\times r}, S_i \in \mathbb R^{m\times n_i}.$
\For {$t=0$ to $T-1$}
\State Server broadcasts $U^{(t)}$ to all clients.
\For {each client $i\in [E]$ (concurrently)}
\State Set $U_i^{(0)}$ as $U^{(t)}$ and $V_i^{(0)},S_i^{(0)}$ as $V_i^{(K)},S_i^{(K)}$ from the last epoch.
\For {$k=0$ to $K-1$}
\vspace{-5pt}
\begin{align}
    & (V_i^{(k+1)}, S_i^{(k+1)}) \leftarrow \arg\min_{V,S} \frac{1}{2}\|U^{(k)} V^T + S - M_i\|^2_F + \frac{\rho}{2}\|V\|_F^2 + \lambda \|S\|_1\label{eq:VS_arg}\\
    & U_i^{(k+1)} \leftarrow U^{(k)} - \eta \nabla_U \mathcal L_i(U^{(k)},V_i^{(K)}, S_i^{(k)})\label{eq:U_local}
\end{align}
\vspace{-10pt}
\EndFor
\State Send back the updated $U_i^{(t)} = U_i^{(K)}$ to the server
\EndFor
\State Server aggregates all $U_i$ by average:

\vspace{-10pt}
\begin{equation}
    U^{(t+1)} \leftarrow \frac{\sum_{i=1}^E U_i^{(t)}}{E} \label{eq:U_agg}
\end{equation}
\vspace{-10pt}
\EndFor
\State {$L_i = U^{(T)} {V_i^{(K)}}^T$ and $S_i = S_i^{(K)}$}
\State \Return {Recovered matrices $\{(L_i,S_i)\}_{i \in \mathcal I_{\text{public}}}$}
\end{algorithmic}
\end{algorithm}

Here, we present a distributed consensus-factorization based RPCA algorithm \name to solve the problem defined in \cref{subsec:DRPCA}. 
We summarize our \name algorithm in \cref{algorithm1}. 
This algorithm uses the nonconvex objective function \cref{eq:noncovex_problem} to avoid using nuclear norm. We claim that this objective function is perfectly separable for each client, making it suitable for the distributed optimization.

We define local objective functions for each $i\in [E]$:
\begin{equation}
    \tilde{\mathcal L_i}(U_i,V_i,S_i) = \frac{1}{2}\|U_iV_i^T + S_i - M_i\|_F^2 + \frac{\rho}{2}\|V_i\|_F^2 + \lambda \|S_i\|_1. 
\end{equation}
The overall optimization goal \cref{eq:noncovex_problem} can be written as a summation over each client: $\mathcal L(U,V,S) = \sum_{i=1}^E \tilde{\mathcal L_i}(U_i,V_i,S_i) + \frac{\rho}{2}\|U\|_F^2$. As a result, \cref{eq:noncovex_problem} can be decomposed into multiple subproblems for each client to solve. 

In order to enforce $\rank([L_1\ L_2\ \dots\ L_E])\leq r$, we require each client to reach a consensus on their left matrix $U_i$, i.e., every matrix $U_i$ must be identical. Therefore, we absorb the $\frac{\rho}{2}\|U\|_F^2$ term into each $\tilde{ \mathcal L_i}(U_i,V_i,S_i)$, so the local objective under consensus is
\begin{equation}
    \mathcal L_{i}(U,V_i,S_i) = \tilde{\mathcal L_i}(U, V_i,S_i) + \frac{n_i\rho}{2n} \|U\|_F^2.
\end{equation}
The problem can thus be reformulated into finding a solution for
\begin{equation}
    U^* \in \arg\min_U g(U) = \arg\min_U \sum_{i=1}^E g_i(U)  \label{eq:U_goal}
\end{equation}
where
\begin{align}
    g_i(U) &= \inf_{V_i,S_i} \mathcal L_i(U, V_i, S_i)\nonumber\\
    &= \inf_{V_i,S_i} \left(\frac{1}{2}\|U V_i^T + S_i - M_i\|_F^2 + \frac{\rho}{2}(\|V_i\|_F^2 + \frac{n_i}{n}\|U\|_F^2)+\lambda \|S_i\|_1\right)
    \label{eq:VS}
\end{align}

\cref{eq:VS_arg} is a minimization problem of a convex function and the details of solving it is explained later. \cref{eq:U_local} updates the matrix $U$ locally based on the surrogate optimal solution $(V_i^*,S_i^*)$. As we will see in (\cref{subsec:converge} (\cref{lemma:2}), \cref{eq:U_local} executes one step of local gradient descent for $U$ optimizing the local objective $g_i(U)$, when the solution $(V_i^*, S_i^*)$ is exact.

Finally, \cref{eq:U_agg} aggregates the outputs from remote clients by average. This scheme is known as the FedAvg \cite{fedavg} algorithm in federated learning.
When $K = 1$, \cref{eq:U_agg} is equivalent to performing exactly one step of the global gradient descent following \cref{eq:U_goal}. However, in practise, the communication cost among remote clients is non-negligible. Setting $K > 1$ allows clients to run local gradient descent for multiple steps, reducing the communication overheads caused by synchronization. Moreover, it has been shown that choosing either a diminishing learning rate or a carefully designed fixed learning rate $\eta = O(\frac{1}{\sqrt{KT}})$ guarantees the convergence of FedAvg algorithm \cite{Li2020On,Fedconverge}.

\textbf{Details of solving \cref{eq:VS_arg}.} Here we elaborate the details for optimizing the convex function \cref{eq:VS_arg}. We claim that the solution for the convex local optimization problem

% We propose to solve this equation by the alternating direction method.
% \begin{align}
%   & V_i^{(p+1)} \leftarrow \arg\min_{V\in \mathbb R^{n_i\times r}} \frac{1}{2}\|U^{(k)}V^T + S_i^{(p)} - M_i\|_F^2 + \frac{\rho}{2}\|V\|_F^2.\label{eq:V}\\
%   & S_i^{(p+1)} \leftarrow \arg\min_{S\in \mathbb R^{m\times n_i}} \frac{1}{2}\|U^{(k)}{V_i^{(p)}}^T + S - M_i\|_F^2 +  \lambda \|S\|_1 \label{eq:S}
% \end{align}
% Both updates described in \cref{eq:V,eq:S} have closed-form solutions. \cref{eq:V} is equivalent to \todo{Add complexity analysis in sec 3}
% \begin{equation}
% \label{eq:Vupdate}
%     {V_i^{(k+1)}}^T = (U^TU + \rho I)^{-1} U^T (M_i - S_i^{(k)}),
% \end{equation}
% while \cref{eq:S} is equivalent to 
% \begin{equation}
% \label{eq:Supdate}
%     S_i^{(k+1)} = \mathrm{sign}(M_i - U{V_i^{(k)}}^T)\cdot \max(|M_i - U{V_i^{(k)}}^T| - \lambda, 0)
% \end{equation}

\begin{equation}
\label{eq:local_obj}
    \{V_i^*,S_i^*\} = \arg\min_{V_i,S_i} \frac{1}{2}\|UV_i^T + S_i - M_i\|^2_F + \frac{\rho}{2}\|V_i\|^2_F + \lambda \|S_i\|_1
\end{equation}
is unique. This is because $(V_i, S_i)$ is the solution for \cref{eq:local_obj} only if
\begin{align}
    &(U_i^T U_i + \rho I) V^T V = U_i^T (M_i-S_i) V_i\\
    & S_i = \mathrm{sign}(M_i - U_i V_i^T)\cdot \max(|M_i - U_iV_i^T| - \lambda, 0)\label{eq:optS}
\end{align}

Bringing \cref{eq:optS} back to \cref{eq:local_obj} yields:
\begin{align}
    V_i^* &\in \arg\min_{V_i\in \mathbb R^{n_i\times r}} \left[\frac{\rho}{2}\|V_i\|_F^2 + \min_{S_i\in \mathbb R^{m\times n_i}} \left(\frac{1}{2}\|UV_i^T + S_i - M_i \|_F^2 + \lambda \|S_i\|_1\right)\right]\nonumber \\
    &=\arg\min_{V_i\in \mathbb R^{n_i\times r}} \left(\frac{\rho}{2}\|V_i\|_F^2 + H_{\lambda}(M_i - UV_i^T)\right). \label{eq:cancelV}
\end{align}
where $H_\lambda(\cdot)$ is the Huber loss, as defined in \cref{appendix:defer}. We denote the inner objective by $h(V_i) = \frac{\rho}{2}\|V_i\|_F^2 + H_\lambda (M_i - UV_i^T)$. We show by \cref{lemma:VS_converge} in \cref{subsec:converge} that $h(V_i)$ is $\rho$-strongly convex, which means the solution for \cref{eq:cancelV} is unique.
Moreover, \cref{lemma:VS_converge} guarantees a linear convergence for applying gradient descent on $V_i$ to optimize $h(V_i)$. As a result, the convex problem in \cref{eq:VS_arg} can be solved efficiently.

% \begin{align}
%     g_i(U) &= \frac{n_i\rho}{2n} \|U\|_F^2 + \min_{V_i\in \mathbb R^{n_i\times r}} \left[\frac{\rho}{2}\|V_i\|_F^2 + \min_{S_i\in \mathbb R^{m\times n_i}} \left(\frac{1}{2}\|UV_i^T + S_i - M_i \|_F^2 + \lambda \|S_i\|_1\right)\right]\nonumber \\
%     &= \frac{n_i\rho}{2n} \|U\|_F^2 + \min_{V_i\in \mathbb R^{n_i\times r}} \left(\frac{\rho}{2}\|V_i\|_F^2 + H_{\lambda}(M_i - UV_i^T)\right).\label{eq:gi}
% \end{align}

\textbf{Problems with Unknown Exact Rank} Attentive readers may notice that factorization-based algorithms including our \cref{algorithm1} require knowing the exact rank $r$ of the underlying low-rank matrix $L_0$. A more general problem formulation \cite{Sha_2021} considers anticipating an upper bound for the rank of $L_0$, i.e., $\rank(L_0) \leq p$.

To solve this harder problem, \name is slightly modified such that $U\in \mathbb R^{m\times p}$ and $V_i\in \mathbb R^{n_i\times p}$. As long as the incoherence condition \cite{Candes2009} (as explained detailedly in \cref{appendix:defer}) is satisfied, the global minimizer of \cref{eq:noncovex_problem} is still guaranteed to be the exact recovery, because of the property of nuclear norm: $\|L\|_* = \inf_{U,V} \frac{1}{2}\{\|U\|_F^2 + \|V\|_F^2: UV^T = L\}$. We also confirm this statement numerically in \cref{sec:exp}.

\textbf{Privacy Preserving.}
We claim that \name also works for privacy critical scenarios. It learns a left matrix $U$ on consensus over all clients, but the private right matrices $V_i$ are kept secret for individuals. As suggested in \cref{algorithm1}, \name reveals $L_i$ only for public data $i\in \mathcal I_{\text{public}}$ and keeps $M_i$ secret for $i\in \mathcal I_{\text{private}}$.

\section{Theoretical Analysis}
\label{sec:theoretical}
\subsection{Preliminaries}

We first define some terminologies before going to details of the convergence analyses. 
\begin{definition}[Smoothness]
\label{def:smooth}
Consider a $C^1$-continuous function $f:\mathbb R^d \to \mathbb R$. We call the function \textbf{$L$-smooth} if its derivative is $L$-Lipschitz, i.e.,
\begin{equation}
    \|\nabla f(w_1) - \nabla f(w_2)\|_2 \leq L \|w_1 - w_2\|_2.
\end{equation}
\end{definition}

% \begin{definition}[Polyak-\L ojasiewicz (PL)]
% \label{def:PL}
% We call a $C^1$-continuous function $f:\mathbb R^d \to \mathbb R$ \textbf{$\mu$-Polyak-\L ojasiewicz (PL)}, if it satisfies
% \begin{equation}
%     \frac{1}{2}\|\nabla f(w)\|_2^2 \geq \mu (f(w) - f(w^*)), \forall w\in \mathbb R^d
% \end{equation}
% where $w^* \in \arg\min_w f(w)$.
% \end{definition}

% The $\mu$-PL condition of \cref{def:PL} is a generalized version of $\mu$-strongly convexity, since a $\mu$-strongly convex function must satisfies the $\mu$-PL condition. However, a $\mu$-PL function does not necessarily has convexity, which is thus studied as a tool for analyzing nonconvex functions \cite{PLconverge}.

\begin{definition}[Strongly convex]
\label{def:strongly_convex}
We call a $C^1$-continuous function $f:\mathbb R^d\to \mathbb R$ \textbf{$\mu$-strongly convex}, if $f(w) - \frac{\mu}{2}\|w\|_2^2$ is a convex function.
\end{definition}

\subsection{Convergence Analysis}
\label{subsec:converge}
In this section, we analyze the convergence rate of our distributed RPCA algorithm. 
The optimization variables are assumed to be bounded during the whole training.

\begin{assumption}[Bounded variables]\label{assum:bound}
During training, all the variables $U,V,S$ are bounded, i.e.,
\begin{equation}
    \|U\|_F \leq C_U, \|V_i\|_F \leq C_V, \|S_i\|_F \leq C_S, \|M_i\|_F \leq C_M, \forall i\in [E].
\end{equation}
\end{assumption}

Based on \cref{assum:bound}, we first state several lemmas regarding the local optimization problems \cref{eq:VS,eq:cancelV}. The detailed proofs are omitted to \cref{appendix:lemma}.
\begin{lemma}
\label{lemma:VS_converge}
The objective function $h(V_i)$ is $(\rho + C_U^2)$-smooth and $\rho$-strongly convex.
\end{lemma}

\begin{lemma}
\label{lemma:2}
    The objective function $g_i(U)$ is differentiable and for any $U$,
    \begin{equation}
        \nabla_U g_i(U) = \nabla_U \mathcal L_i(U,V_i^*, S_i^*).
    \end{equation}
\end{lemma}

\begin{lemma}
\label{lemma:smooth}
The local objective function $g_i(U)$ defined in \cref{eq:VS} is $L$-smooth, where
\begin{equation}
    L = r + C_V^2 + \frac{4C_S + 12C_V + 4C_M}{\rho} C_VC_U.
\end{equation}
\end{lemma}

\begin{lemma}
\label{lemma:bounded_grad}
The local objective function $g_i(U)$ has bounded gradient
\begin{equation}
    \|\nabla_U g_i(U)\|_F \leq C_U \sqrt{\frac{mn_i^2\rho^2}{n^2} + m^2 n_i \lambda^2}.
\end{equation}
\end{lemma}

With the help of \cref{lemma:smooth,lemma:bounded_grad}, we show that our \name algorithm converges with a rate of $O(\frac{1}{\sqrt{KT}})$

\begin{theorem}
\label{converge1}
If the learning rate $\eta < \frac{1}{L}$, the average squared gradient of \name converges by
\begin{equation}
    \frac{1}{T} \sum_{t=0}^{T-1} \|\nabla_U g(U^{(t)})\|_F^2 \leq \frac{2(g_0 - g^*)}{\eta K T} + 4\eta^2 K^2 C_U^2 L^2 (m\rho^2 + m^2 n \lambda^2)
\end{equation}
\end{theorem}
\emph{Proof Sketch.} The proof of \cref{converge1} is straight-forward, given the sufficient literature in distributed learning of analyzing the FedAvg algorithm \cite{Li2020On,Yu2019ParallelRS}. We defer the detailed proof to \cref{appendix:thm1}, in which we leverage the conclusions from \cref{lemma:smooth,lemma:bounded_grad} to prove the theorem.

\textbf{Remark.} We choose $\eta = \frac{c}{\sqrt{KT}}$ in \cref{converge1}, so that the norm of gradient converges to zero.
\begin{equation}
    \frac{1}{T} \sum_{t=0}^{T-1} \|\nabla_U g(U^{(t)})\|_F^2\leq \frac{2(g_0 - g^*)}{c\sqrt{KT}} + \frac{4c^2 K C_U^2 L^2 (m\rho^2 + m^2 n \lambda^2)}{T}.
\end{equation}

Though \cref{converge1} shows the convergence of the \name algorithm, it does not implies any clues to the stationary point reached.

\subsection{Hyperparameter Analysis}

As we have stated in \cref{subsec:converge}, \name optimizes over a nonconvex objective function and may not converge to a global optimal solution. Here we first state a necessary condition for the hyperparameters $\rho, \lambda$ of finding the exact solution.

\begin{theorem}
\name finds a global optimal solution only if
\begin{equation}
    \rho^2 \leq \lambda^2 mn.
\end{equation}
\end{theorem}
\emph{Proof Sketch.} We prove this theorem by showing when $\rho^2 > \lambda^2 mn$, the gradient is always nonzero unless $U = 0$. Therefore, $\rho^2 \leq \lambda^2 mn$ is a necessary condition for finding the exact optimal solution.

\subsection{Complexity Analysis}

\name exploits the superiority of distributed computation on large-scale problems by coordinating remote devices through limited communications. In this section, we analyze the complexity of \name in two aspects: individual computation cost and the inter-client communication overhead.

\textbf{Computation cost.}
In each local iteration, a remote client first finds the optimal solution for \cref{eq:VS}. The computation of gradient for $h(V_i)$ takes $O(m n_i +r m n_i ) = O(rmn_i)$ operations. As stated in previous sections, the inner objective function $h(V_i)$ converges linearly. Therefore, it takes $O\left(rmn_i \log (\frac{1}{\epsilon})\right)$ time to converge to an $\epsilon$-optimal solution. 
The client then executes one step of gradient descent on $U_i$, in which the gradient can be computed in $O(mr^2+n_ir^2)$ operations. In conclusion, each local iteration takes $O\left(mr \max(r,n_i\log (\frac{1}{\epsilon}) )\right)$ operations to compute. Moreover, if the data are evenly distributed so that $n_i = O(\frac{n}{E})$, the time complexity of each communication round for each client is
\begin{equation}
    T_{\mathrm{local}} = O\left(Kmr \max(r,\frac{n}{E}\log \frac{1}{\epsilon})\right).
\end{equation}

A central server is in charge of aggregating $E$ updated left matrices $\{U_i\}_{i\in [E]}$ by average. The amount of its work load is
\begin{equation}
    T_{\mathrm{server}} = O\left(mr\log E \right).
\end{equation}

\textbf{Communication cost.}
In each communication round, the server broadcasts a matrix of size $m\times r$ to all clients, while each client sends an updated matrix of the same size back to the server at the end of this round. Therefore, the total communication overhead in each round is
\begin{equation}
    T_{\text{comm}} = 2Emr.
\end{equation}

Regarding both computation and communication costs, the best configuration for the number of clients is $E = O(\sqrt{n})$, so that the overall time cost for each round is
\begin{equation}
    T_0 = O(Kmr \max(r,\sqrt{n} \log \frac{1}{\epsilon})).
\end{equation}

% \todo{Estimate runtime for other RPCA algorithms for comparison.}
In common real world scenarios when numerous remote agents run a low-rank approximation algorithm jointly, the local data volume $n_i$ is typically bounded. The individual computation cost and communication overhead remain constant as the number of clients increases. Therefore, \name is scalable in terms of the data scale $n$ and the number of clients $E$. It is particularly superior when $m$ is fixed and is much smaller than $n$. We claim that this case is common for distributed deep learning as $m$ corresponds the data dimension or the number of extracted features and $n$ stands for the total size of datasets.

\section{Experimental Evaluation}
\label{sec:exp}

In this section, we present the experimental results for \name. The experimental setups and evaluation metrics are first introduced in \cref{exp:setup}. We then compare \name with other RPCA algorithms on different problem scales and show xxx in \cref{exp:main}. In \cref{exp:ablation} we conduct experiments on different hyperparameter choices for \name and perform ablation studies.

\subsection{Experimental Setup}
\label{exp:setup}

\textbf{Problem Generation.} The RPCA problems are generated by the following scheme for all experiments. We first randomly sample a ground truth low rank matrix $L^*$ by $L_0 = U_0V_0^T$, where $U_0\in \mathbb R^{m\times r}$ and $V_0 \in \mathbb R^{n\times r}$ are random Gaussian matrix whose entries are sampled from the standard Gaussian distribution $\mathcal N(0,1)$. We then sample a sparse matrix $S_0$ with $smn$ random nonzero entries, $0<s<1$. Each entry of $S_0$ is sampled from $\{-\sqrt{mn},0,\sqrt{mn}\}$.

\textbf{Evaluation Metric.} We evaluate the recovery of the low rank matrix by a relative error rate for $L$ and $S$.
\begin{equation}
    \mathrm{err} = \frac{\|L- L_0\|_F^2 + \|S-S_0\|_F^2}{\|L_0\|_F^2 + \|S_0\|_F^2}.
\end{equation}

\textbf{Implementation.} \name uses a server-client distributed paradigm and should be implemented distributedly in reality. However, we simulate the \name algorithm on a single device in our experiments instead for clarity. We run each local program sequentially and only allow communications when the server synchronizes the programs. For comparison, we implement two common centralized algorithms based on convex relaxation,  APGM\cite{Lin2009FastCO} and ALM\cite{Goldfarb2013FastAL}.

\subsection{Main Results}
\label{exp:main}

\begin{figure}[tb]
    \centering
    \includegraphics[width=0.8\textwidth]{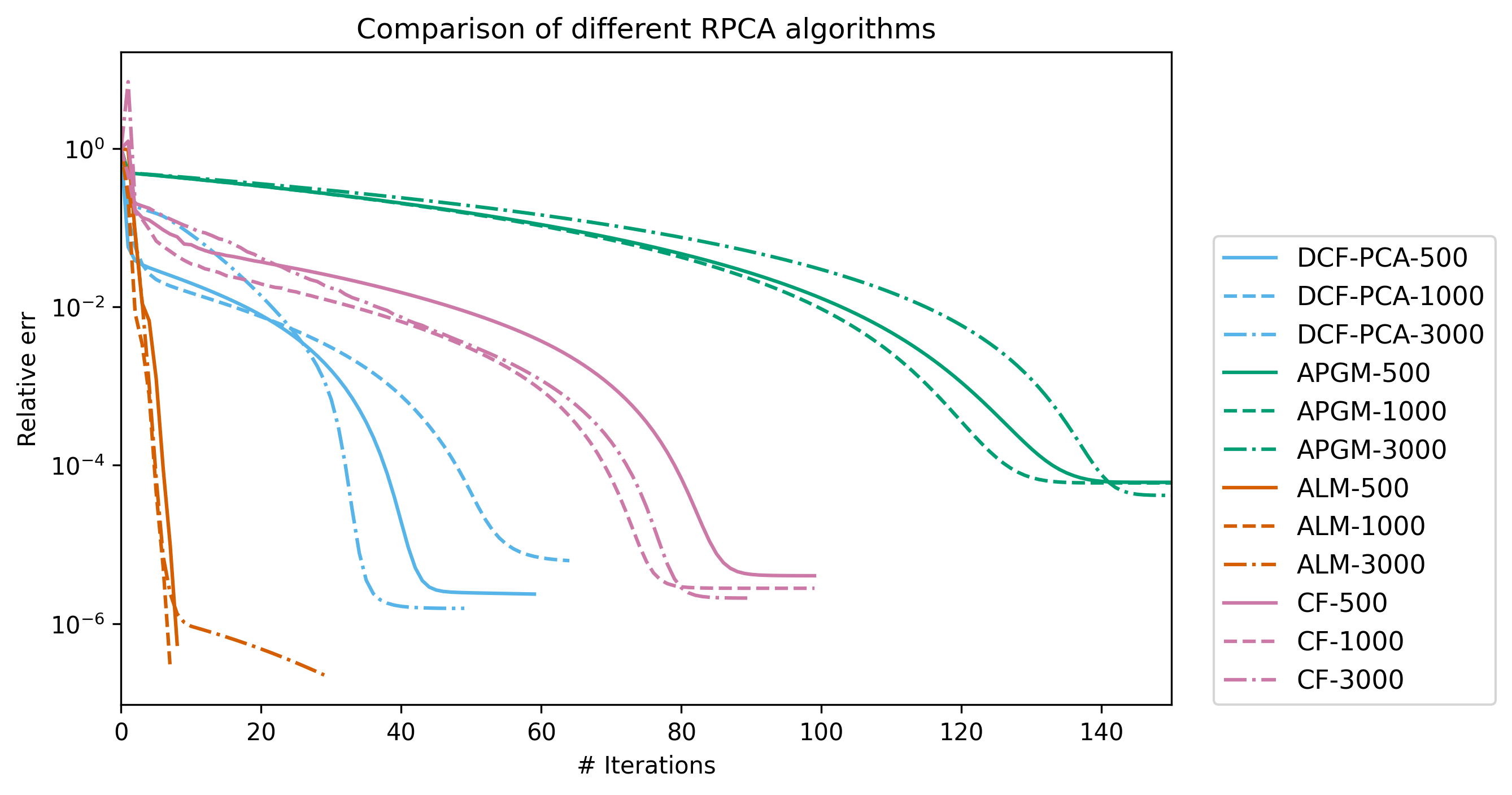}
    % \begin{subfigure}[b]{0.45\linewidth}
    % \centering
    % \includegraphics[width=\textwidth]{figures/M500.png}
    % \caption{$m=n=500$}
    % \end{subfigure}
    % \begin{subfigure}[b]{0.45\linewidth}
    % \centering
    % \includegraphics[width=\textwidth]{figures/M1000.png}
    % \caption{$m=n=1000$}
    % \end{subfigure}
    \caption{Comparison on the convergence of different algorithms for the synthetic RPCA problems of different scales ($m=n=500,1000,3000$).}
    \label{fig:main}
\end{figure}

\textbf{Exact rank recovery}
\cref{fig:main} compares the performance of different algorithms in solving the RPCA problems. We let $m=n$ for all experiments and choose $r = 0.05 n, s = 0.05$ for generating the target matrices.
We also report the performance for the centralized version of \name as a baseline, which is denoted as CF-PCA in \cref{fig:main}. Among all algorithms compared in \cref{fig:main}, only \name runs distributedly. As a result, \name costs much less computation time than its centralized counterpart CF-PCA. We note that different learning rates are used for \name and CF-PCA. The distributed \name needs small learning rate for keeping consensus on the matrix $U$, while the single-thread CF-PCA makes use of a larger learning rate for efficiency. For all experiments, we use decaying learning rate $\eta = O(\frac{\eta_0}{\sqrt{t}})$.

\begin{figure}[tb]
\centering
    \includegraphics[width=0.6\textwidth]{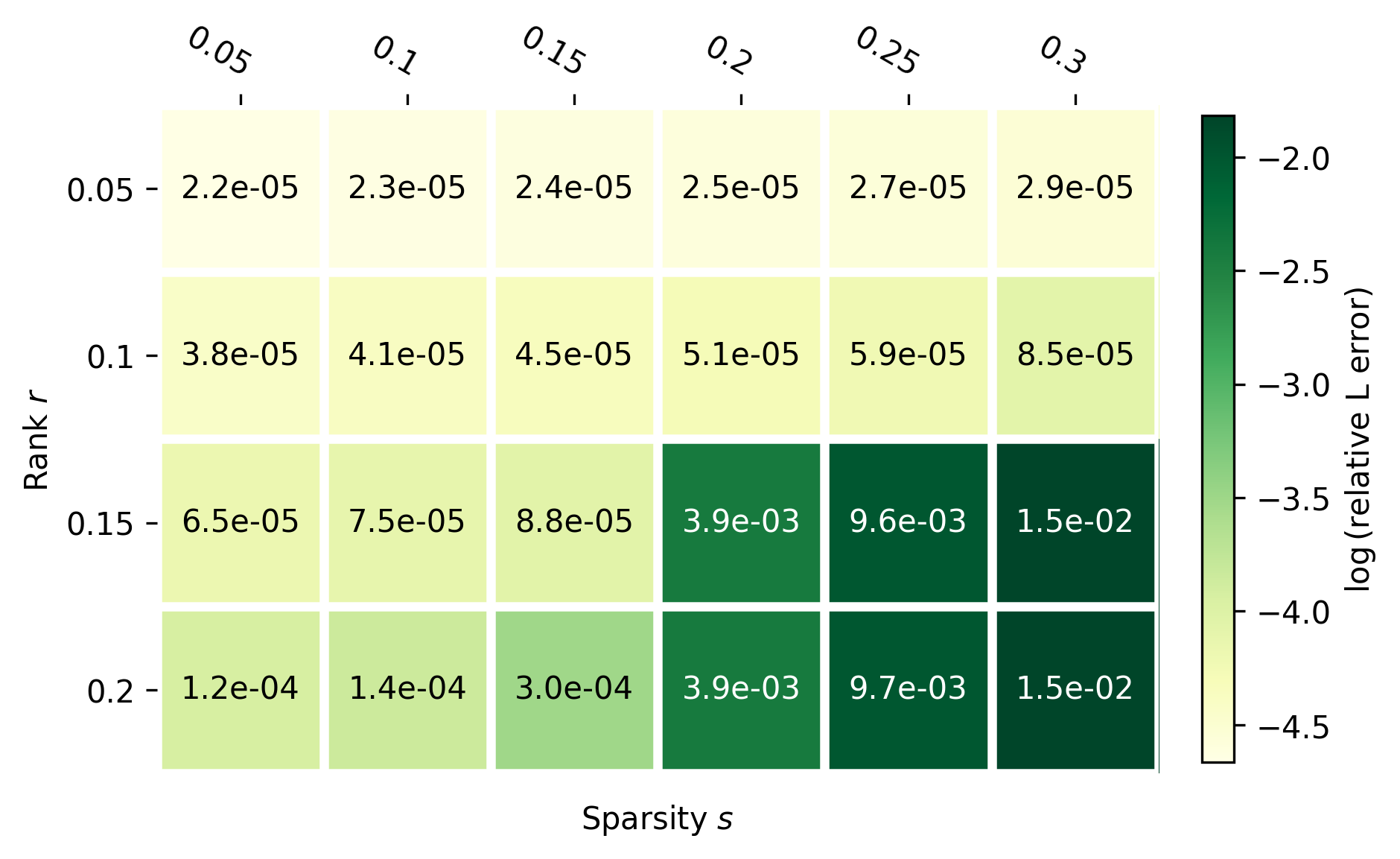}
    \caption{Relative error of the recovered matrices under different sparsity and low-rank levels (m=n=500).}
    \label{fig:sparse}
\end{figure}

We also test the performance of \name on matrices with different levels of sparsity and low rank. In \cref{fig:sparse} we report the relative error of the recovered matrices for different problem configurations, including $s \in [0.05, 0.3]$ and $r\in [0.05n,0.2n]$. We run \name with less than 50 iterations with $K = 2$ and initial learning rate $\eta_0 = 0.05$. A distinctive limit occurs at $r\approx 0.15n$ and $s\approx 0.2$. Any target matrix with larger ground truth rank $r$ and larger sparsity $s$ than the limit cannot be recovered correctly.

% \todo{Evaluate harder cases and report success rates.}

\textbf{Upper-bound rank recovery}

\begin{figure}[t]
    \centering
    \includegraphics[width=0.6\textwidth]{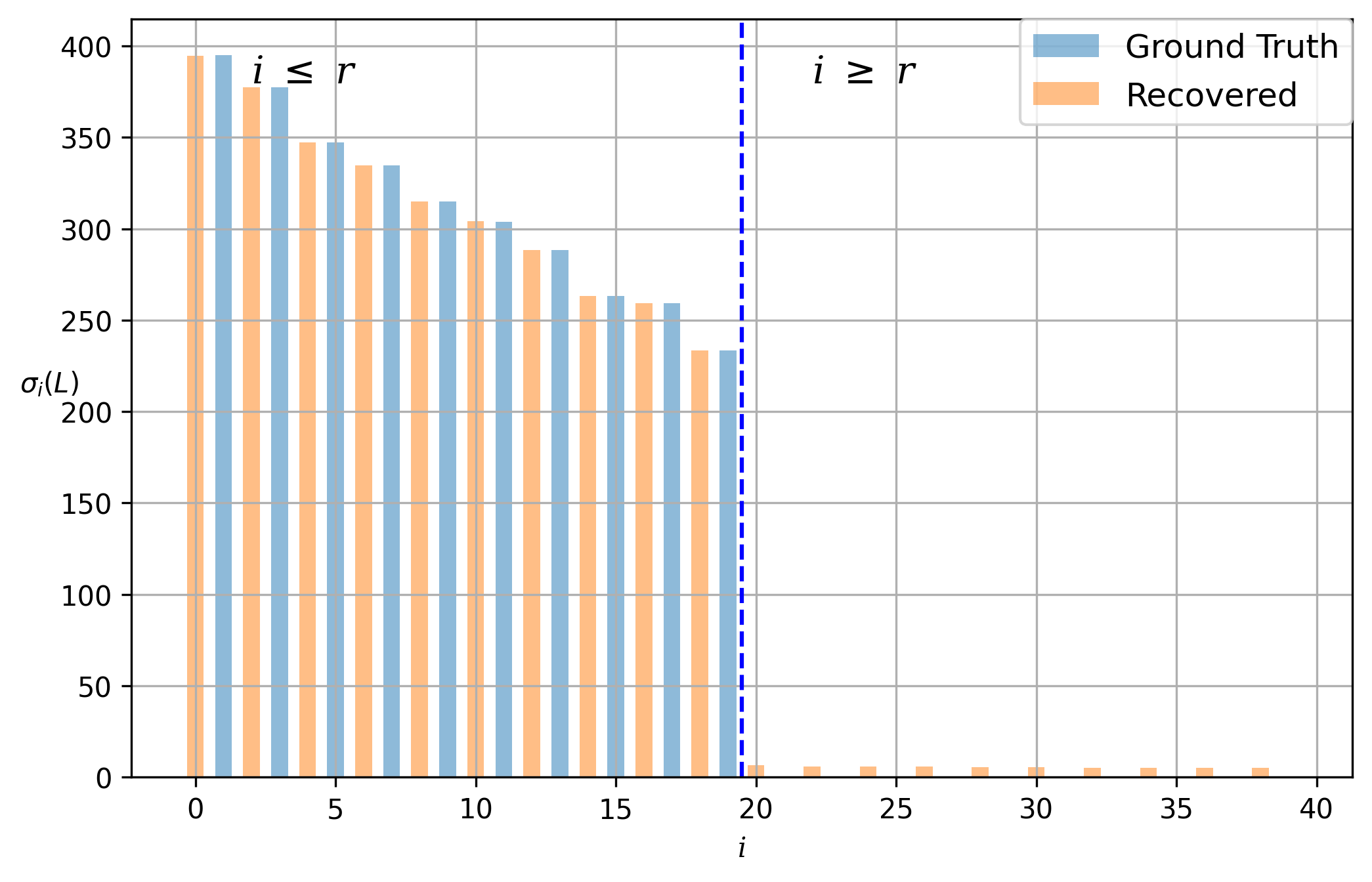}
    \caption{Comparison of the singular values between the recovered matrix and the ground truth matrix.}
    \label{fig:singular_values}
\end{figure}

Here we present the evaluation results for \name without anticipating the exact rank of $L$, but only with an upper bound $r \leq p$ on the rank. \cref{fig:singular_values} compares the singular values of the recovered matrix $L^{(T)}$ with the original low-rank matrix $L_0$, when $M=N=200$, $r=0.05 n, s = 0.05$ and $p = 0.1 n$. It shows that the recovered matrix with $p = 2r$ successfully approximate the ground truth matrix of rank $r$, as $\sigma_{r+1}(L^{(T)}) / \sigma_r (L^{(T)})$ is small.
Quantitatively, we report the relative singular value error: $\frac{\max|\sigma_{i}(L^{(T)})-\sigma_i(L_0)|}{\sigma_r(L_0)}$ in \cref{table:singular} for various problem scales.

\begin{table}[t]
    \caption{Relative singular value errors of \name for different problem scales.}
    \label{table:singular}
    \centering
    \begin{tabular}{lllc}
    \toprule
    n & r & p & $\frac{\max|\sigma_{i}(L^{(T)})-\sigma_i(L_0)|}{\sigma_r(L_0)}$ \\
    \midrule
    200 & 10 & 20 & 0.0286\\
    500 & 25 & 50 & 0.0326\\
    1000 & 50 & 100 & 0.0398\\
    5000 & 250 & 500 & 0.1127\\
    \bottomrule
    \end{tabular}
\end{table}

\subsection{Ablation Studies}
\label{exp:ablation}

\textbf{Number of local iterations}

We study the influence of different $K$ values in \name to the convergence rate. As $K$ denotes the number of local iterations per communication round, it reflects the level of asynchronization among remote clients. When $K = 1$, the \name algorithm is fully synchronized and $U$ descends exactly along the gradient of the global objective.

\begin{figure}[tb]
    \centering
    \includegraphics[width=0.6\textwidth]{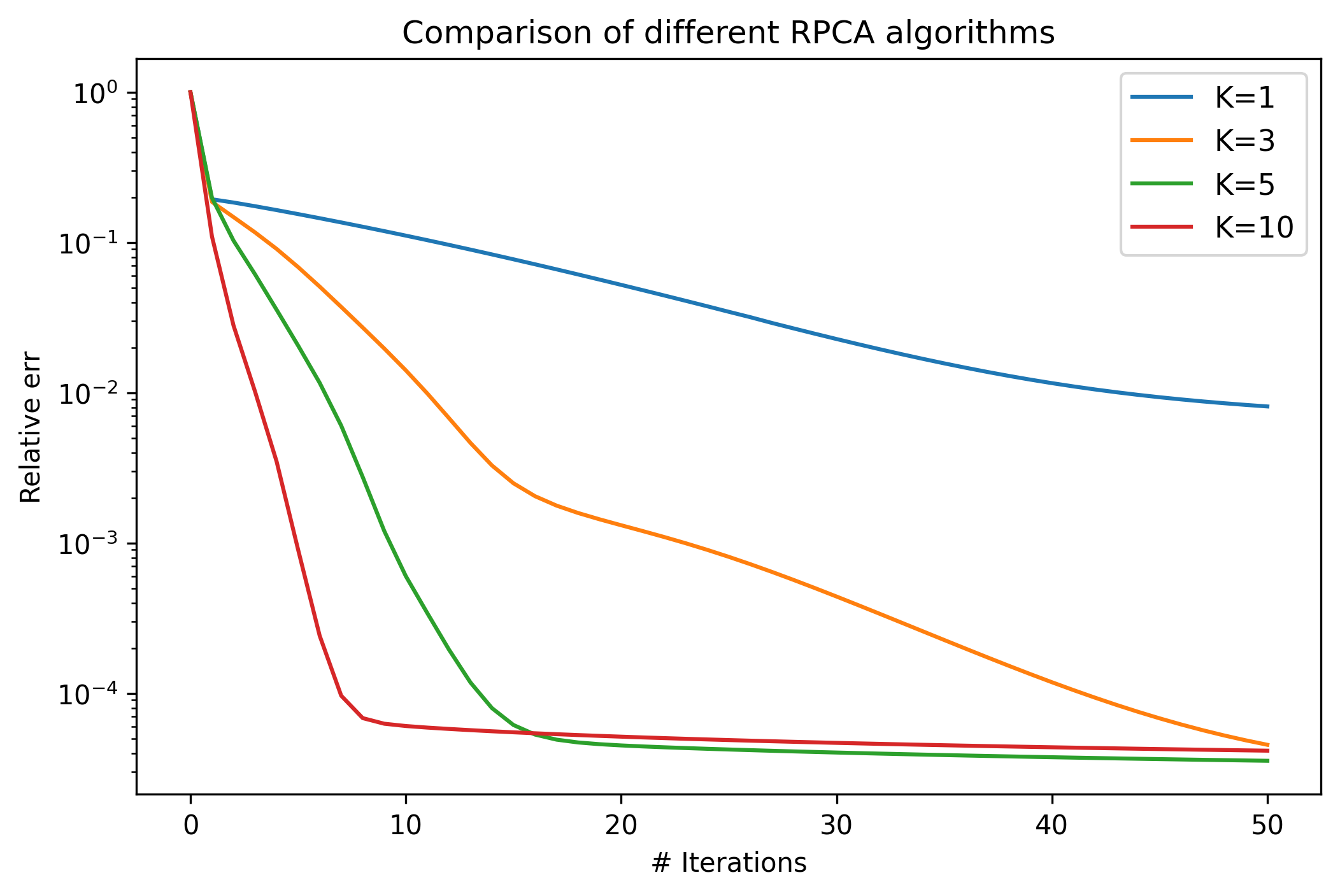}
    \caption{Comparison on different numbers of local iterations $K$. It only takes 8 iterations for \name with $K=10$ to converge; while $K=1$ converges much slower.}
    \label{fig:1}
\end{figure}

\cref{fig:1} shows the convergence of \name with the same learning rate $\eta = 0.01$ and the number of clients $E = 10$, but with different values of $K$. Our algorithm converges remarkably faster as $K$ increases, but also suffers from a slightly larger error floor at the same time.

\section{Conclusions}

In this work we formalize the problem of distributed robust principal analysis and present a distributed consensus-factorization algorithm (\name) that can run distributedly in remote devices. Our algorithm does not require sharing the target matrices among remote clients and only consumes limited communication. The nice properties of \name help preserving private data of clients and speeding up the computation of large-scale RPCA problems. \name is particularly effective for scenarios when $n$ represents the total number of items in a large distributed dataset while $m$ represents the data dimension or the extracted features dimension in deep learning applications, since $n \gg m$ is usually observed in these cases.
We show the convergence of our algorithm both theoretically and numerically. \name is also shown not sensitive to the choices of the number of local iterations $K$ and predefined rank upper bound $p$.\\

% \textbf{Claim:} This project has not been used for other purposes except for the Term project.
\clearpage

% \begin{lemma}
% \label{lemma:PL}
% The global objective function $g(U) = \sum_{i=1}^E g_i(U)$ is $\mu$-PL, where
% \begin{equation}
%     \mu = 
% \end{equation}
% \end{lemma}

% \bibliographystyle{plain}
\bibliography{reference}

\newpage
\appendix
% \part*{} % Start the appendix part
% \DoToC
% \newpage

\section{Omitted Preliminaries}
\label{appendix:defer}

\subsection{Additional Lemmas}

\textbf{Danskin's Theorem}

Here we recall a finer version of the Danskin's Theorem as stated in \cite{Bonnans1998}.
\begin{lemma}[Theorem 4.1 \cite{Bonnans1998}]
\label{lemma:danskin}
Let $f:\mathbb R^{a} \times \mathbb R^{b}\to \mathbb R$. Suppose $f(x,\cdot)$ is differentiable and that $f(x,u), \nabla_u f(x,u)$ are continuous. Let $\Phi \subseteq \mathbb R^{a}$ be a compact subset. Then $g(u) = \min_{x\in \Phi} f(x,u)$ is directionally differentiable. Moreover, when $f(\cdot, u)$ has a unique minimizer $x^*$ over $\Phi$,
\begin{equation}
    \nabla_u g(u) = \nabla_u f(x,u).
\end{equation}
\end{lemma}

\subsection{Huber Loss}
The huber loss $H_\lambda(\cdot)$ is defined as
\begin{equation}
    H_\lambda(x) = \begin{cases} 
    - \lambda x - \frac{\lambda^2}{2}, \quad \text{if}\ x < -\lambda.\\
    \frac{x^2}{2}, \quad \text{if}\ -\lambda \leq x \leq \lambda \\
    \lambda x - \frac{\lambda^2}{2},\quad \text{if}\ x > \lambda.
    \end{cases}
\end{equation}

One can quickly check that $H_\lambda(x)$ is convex and differentiable. Furthermore, if $X\in \mathbb R^{m\times n}$ is a matrix, we define $H_\lambda(X) = \sum_{i=1}^m \sum_{j=1}^n H_\lambda(X_{ij})$.

\subsection{Incoherence condition}

The general problem formulation for robust PCA is ill-posed. It assumes the target matrix $M$ can be decomposed to a low-rank matrix $L$ and a sparse matrix $S$, but there might be multiple possible solutions to it. Moreover, when $L$ is also sparse or $S$ has low rank, the recovery process is meaningless. 
\cite{Candes2009} presents the incoherence conditions for RPCA problems that guarantees $L$ to be dense enough and $S$ to have larger enough rank, so that the recovery by optimizing convex relaxed objective \cref{eq:convex1} is exact. The incoherence condition is stated as
\begin{align}
    & \max_i \|U^T e_i\|^2 \leq \frac{\delta r}{m}, \max_i \|V^T e_i\|^2 \leq \frac{\delta r}{n}\\
    & \|UV^T\|_\infty \leq \sqrt{\frac{\delta r}{mn}}.
\end{align}
where $L = U\Sigma V^T$ is the singular value decomposition of $L$.
Later, it is shown that under this assumption, \cref{eq:convex2} also recovers $L$ with high probability.

\section{Proofs}

\subsection{Deferred Proofs of Lemmas}
\label{appendix:lemma}

\textbf{Lemma 1.}
\begin{proof}
Note that the Huber loss function $H_{\lambda}(\cdot)$ is differentiable, so
\begin{equation}
    \nabla_{V_i} h(V_i) = \rho V_i + H_{\lambda}^\prime(M_i - UV_i^T)^T U.
\end{equation}
For any $V_1, V_2\in \mathbb R^{n_i\times r}$, we have
\begin{align}
    \|\nabla h(V_1) - \nabla h(V_2)\|_F &= \|\rho (V_1 - V_2) + U\left(H_{\lambda}^\prime (M_i - UV_1^T) - H_{\lambda}^\prime(M_i - UV_2^\prime)\right)\|_F\\
    &\leq \rho \|V_1 - V_2\|_F + \|U\|_F \|H_{\lambda}^\prime (M_i - UV_1^T) - H_{\lambda}^\prime(M_i - UV_2^T)\|_F\\
    &\leq \rho \|V_1 - V_2\|_F + \|U\|_F \|UV_1^T - UV_2^T\|_F\\
    &\leq (\rho + C_U^2) \|V_1 - V_2\|_F.
\end{align}
so $h(\cdot)$ is $(\rho + C_U^2)$-smooth. On the other hand, $f(V_i,S_i) =  \frac{1}{2}\|UV_i^T + S_i - M_i\|_F^2 + \lambda \|S_i\|_1$ is convex, so
\begin{equation}
    h(V_i) = \frac{\rho}{2}\|V_i\|_F^2 + \min_{S_i} f(V_i,S_i)
\end{equation}
is $\rho$-strongly convex.
\end{proof}

\textbf{Lemma 2.}
\begin{proof}
Recall that the local objective function is defined as
$$
\mathcal L_i(U, V_i,S_i) = \frac{1}{2}\|UV_i^T + S_i - M_i\|^2_F +  \frac{\rho}{2}(\frac{n_i}{n} \|U\|_F^2 + \|V_i\|_F^2) + \lambda \|S_i\|_1.
$$

It is clear that $\mathcal L(\cdot, V_i,S_i)$ is differentiable. Moreover, both $\mathcal L$ and $\nabla_U L(U,V_i,S_i) = \frac{\rho n_i}{n}U + (UV_i^T + S_i - M)V_i$ are continuous. We also know that for any fixed $U$, $\arg\min_{V_i,S_i}\mathcal L(U,V_i,S_i)$ is unique, since $h(V_i)$ is $\rho$-strongly convex. Therefore, according to \cref{lemma:danskin}, $g_i(U) = \nabla_U \nabla_U \mathcal L(U,V_i,S_i)$.

\end{proof}

\textbf{Lemma 3.} 
\begin{proof}
Let $V_i^*, S_i^* = \arg\min_{V_i,S_i} \mathcal L(U,V_i, S_i)$
\begin{align}
    \nabla_U g_i(U) &= \nabla_U \min_{V_i,S_i} \mathcal L(U,V_i,S_i)\\
    &= \nabla_U \mathcal L(U,V_i^*,S_i^*)
\end{align}

Note that $\mathcal L(U,V,S)$ is strongly convex in terms of $V$ and $S$, we have
\begin{equation}
    \mathcal L_i(U,V_{i}^\prime, S_{i}^\prime) - \mathcal L_i(U, V_{i}, S_{i}) \geq \frac{\rho}{2}\|V^\prime - V\|_F^2 + \frac{1}{2} \|S^\prime - S\|_F^2.
\end{equation}
Similarly,
\begin{equation}
    \mathcal L_i(U^\prime,V_{i}, S_{i}) - \mathcal L_i(U^\prime, V_{i}^\prime, S_{i}^\prime) \geq \frac{\rho}{2}\|V^\prime - V\|_F^2 + \frac{1}{2} \|S^\prime - S\|_F^2.
\end{equation}

On the other hand, 
\begin{equation}
    f(V_i,S_i) = \mathcal L(U^\prime,V_i,S_i) - \mathcal L(U, V_i, S_i)
\end{equation}
is Lipschitz in terms of $V_i$ and $S_i$. This is because
\begin{align}
    &\|\frac{\partial f}{\partial V_i}\|_F = \|V_i ({U^\prime}^TU^\prime - U^T U)\|_F \leq C_V \|{U^\prime}^T U^\prime - U^T U\|_F\leq 2C_VC_U \|U^\prime - U\|_F \\
    &\|\frac{\partial f}{\partial S_i}\|_F = \|(U^\prime - U)V_i\|_F \leq C_V \|U^\prime - U\|_F.
\end{align}

Therefore,
\begin{equation}
    f(V_i^\prime, S_i^\prime) - f(V_i, S_i) \leq 2C_VC_U (\|V_i^\prime - V_i\|_F + \|S_i^\prime - S_i\|_F).
\end{equation}
and
\begin{equation}
    f(V_i^\prime, S_i^\prime) - f(V_i,S_i)\geq \frac{\rho}{2} (\|V^\prime - V\|_F^2 + \|S^\prime - S\|_F^2).
\end{equation}

Combine these two inequalities, we have
\begin{equation}
    \|V_i^\prime - V_i\|_F + \|S_i^\prime - S_i\|_F \leq \frac{4C_VC_U}{\rho} \|U^\prime - U\|_F.
\end{equation}

Now 
\begin{align}
\|\nabla g_i(U^\prime) - \nabla g_i(U)\|_F &\leq \| U_i^\prime - U \|_F \|{V_i^\prime}^T V_i^\prime + \frac{n_i}{n}I\|_F + \|S^\prime V_i^\prime - SV_i\|_F \nonumber\\
&\hspace{7em} +\|M_i (V_i^\prime - V_i)\|_F + \|U\|_F \|{V_i^\prime}^T V_i^\prime - V_i^T V_i\|_F\\
&\hspace{-3em}\leq \|U^\prime - U\|_F \left(\frac{rn_i}{n} + C_V^2 + 4\frac{(C_S+C_V + \|M_i\|_F)C_VC_U}{\rho} + \frac{8C_V^2C_U}{\rho}\right)\\
&\hspace{-3em}\leq\|U^\prime - U\|_F \left(r + C_V^2 + \frac{4C_S + 12C_V + 4C_M}{\rho} C_VC_U\right)
\end{align}
\end{proof}

\textbf{Lemma 4.}
\begin{proof}
The gradient of the local objective $g_i(U)$ is
\begin{align}
    \nabla_U g_i(U) &= \nabla_U \mathcal L(U,V_i^*, S_i^*)\\
    &= (U{V_i^*}^T + S_i^* - M_i)V_i^* + \frac{n_i}{n}U.
    \label{41}
\end{align}

Recall that
\begin{align}
    &(U^TU + \rho I) {V_i^*}^T = U^T(M_i - S_i^*) \label{eq:condition1}\\
    & S_i^* = \mathrm{sign}(M_i - U{V_i^*}^T)\cdot \max(|M_i - U{V_i^*}^T|-\lambda, 0)
\end{align}

Let $\Lambda = M_i - U{V_i^*}^T - S_i^*$, so $\|\Lambda\|_{\infty}\leq \lambda$. Then \cref{eq:condition1} can be rewritten as
\begin{equation}
    \rho V_i^* = \Lambda^T U
\end{equation}

Bringing this back to \cref{41} yields
\begin{equation}
    \nabla_U g_i(U) = -\Lambda V_i^* + \frac{n_i}{n}U = (\frac{n_i}{n}I_m - \Lambda \Lambda^T) U 
\end{equation}

Since $\|\Lambda\Lambda^T\|_F^2 \leq m^2 n \lambda^2$,
\begin{equation}
    \|\nabla_U g_i(U)\|_F \leq \sqrt{(\frac{n_i^2}{n^2}m\rho^2 + m^2 n_i \lambda^2)} \|U\|_F \leq C_U \sqrt{\frac{n_i^2 m\rho^2}{n^2} + m^2 n_i\lambda^2}.
\end{equation}

\end{proof}

\subsection{Proof for \cref{converge1}}
\label{appendix:thm1}
\begin{proof}

We first recall a theorem proved in \cite{Yu2019ParallelRS}:
\begin{lemma}[Yu et al. \cite{Yu2019ParallelRS} Theorem 1]
\label{lemma:3}
Suppose each local objective function is $L$-smooth and has bounded gradient norm $\|\nabla f_i(w)\|_2\leq G$. If each local agent runs local SGD and synchronizes their model weights every $K$ iterations, the FedAvg algorithm converges with
\begin{equation}
    \frac{1}{T}\sum_{i=1}^T \mathbb E[\|\nabla f(w^{(t)})\|^2]\leq \frac{2}{\eta T}(f(w^{(0)} - f^*)) + 4\eta^2 K^2 G^2 L^2 + \frac{L}{N}\eta \sigma.
\end{equation}
where the learning rate $\eta \leq \frac{1}{L}$ and $\sigma$ is the variance of each SGD update.
\end{lemma}

Notice that our \name algorithm has no variance in each local update. We combine \cref{lemma:smooth,lemma:bounded_grad,lemma:3} by $G \leq C_U\sqrt{m\rho^2 + m^2 n\lambda^2}$ and derive the result of \cref{converge1}.

\end{proof}

\end{document}